\documentclass[twoside]{article}
\usepackage{graphicx}
\usepackage{fancyhdr}
\usepackage{multicol}
\usepackage{styl-ap}

\begin{document}
\title{Magnetorotational explosions of core-collapse supernovae}
\author{G.S.Bisnovatyi-Kogan\work{1}, S.G.Moiseenko\work{2}, N.V.Ardeljan\work{3}}
\workplace{Space Research Institute, Profsoyuznaya str. 84/32,
Moscow 117997, Russia, and National Research Nuclear University
"MEPHI", Kashirskoye Shosse, 31, Moscow 115409, Russia
 \next Space Research Institute,
Profsoyuznaya str. 84/32, Moscow 117997, Russia \next Department of
Computational Mathematics and Cybernetics, Moscow State University,
Vorobjevy Gory, Moscow B-234, Russia}
\mainauthor{gkogan@iki.rssi.ru} \maketitle

\begin{abstract}%
Core-collapse supernovae are accompanied  by formation of neutron
 stars. The gravitation energy is transformed into the energy of
 the explosion, observed as SN II, SN Ib,c type supernovae.
 We present results of 2-D MHD simulations, where the source of energy is
 rotation, and magnetic field serves as a "transition belt" for
 the transformation of the rotation energy into the energy of the explosion.
 The toroidal part of the magnetic energy initially grows linearly with time
 due to differential rotation.
 When the twisted toroidal component strongly exceeds the poloidal field,
  magneto-rotational instability develops,
 leading to a drastic acceleration in the growth of magnetic energy.
 Finally, a fast MHD shock is formed, producing a supernova
 explosion. Mildly collimated jet is produced for dipole-like type of the initial field.
 At very high initial magnetic field no MRI development was found.
\end{abstract}

\keywords{Core-collapse supernova - Magnetorotational mechanism -
Numerical modeling}

\begin{multicols}{2}
\section{Introduction}
Supernova is one of the most powerful explosion in the Universe
which releases about $10^{51}$ erg both in radiation and kinetic
energies. SNe explode at the end of evolution of massive stars, with
initial mass larger than $\sim 8M_\odot $.  A thermonuclear
explosion of C-O degenerate core with total disruption of the star
takes place in SN Ia, what happens when initial mass of the star
does not exceed $\sim 12 M_\odot$, when the electrons are degenerate
in the carbon-oxygen core . For larger initial masses the evolution
proceeds until the formation of the iron core, and the star
collapses due to a loss of a hydrodynamic stability.  During the
core collapse and formation of a neutron star, gravitational energy
release $\sim 6\cdot 10^{53}$ erg, is carried away by neutrino. The
first mechanism suggested in \cite{cw66} for the
explanation of of the explosion in a core-collapse SN was connected
with a neutrino deposition. The huge energy flux carried by neutrino
heats the infalling outer layers, reverse the direction of motion,
and leads to formation of the shock wave, producing explosions of SN
II, SN Ib,c types. Later more accurate calculations revealed that
the energy of such explosion is not enough for the explanation of
observations. Many modifications of the neutrino model have been
calculated during  years, but the problem  is not yet clear. The
review of the problem may be found in the book \cite{bk11}.

In recent simulations (see details in \cite{m2012}) we have found that
for extremely strong initial magnetic field $H_0=10^{12}$G a prompt
supernova explosion occurs, and collimated jet is formed in agreement
with \cite{takiwaki2004}. For case when the initial magnetic field
is weaker $H_0=10^9$G we have identified, after the linear growth of the poloidal
magnetic field due to differential rotation, the exponential field
growth due to the magnetorotational instability of Tayler type \cite{tayler}.
We  call the combination of differential rotation with the Tayler type MRI instability,
as magneto-differential-rotation instability (MDRI).
The ejection
due to the explosion is only weakly collimated, while \cite{takiwaki2004}
had obtained a strong collimation in this variant also. In these simulations we considered a
uniform magnetic field along the rotational axis, as the initial
field configuration, similar to \cite{takiwaki2004}.

\section{Magnetorotational mechanism of explosion}

In magnetorotational explosion (MRE) the transformation of the
rotational energy of the neutron star into explosion energy takes
place by means of the magnetic field \cite{bk1970}).
Neutron stars are rotating, and have magnetic fields up to $10^{13}$
Gs, and even more. Often one or two-side ejections are visible. That
indicate to non-spherical form of the SNe explosions. In
differentially rotating new born neutron stars radial magnetic field
is twisted, and magnetic pressure becomes very high, producing MHD
shock by which the rotational energy is transformed to the explosion
energy.

Calculations of MRE have been done in \cite{bkps1976},
using one-dimensional nonstationary equations of magnetic
hydrodynamics, for the case of cylindrical symmetry. The energy
source is supposed to be the rotational energy of the system (the
neutron star, and surrounding envelope). The calculations show that
the envelope splits up during the dynamical evolution of the system,
the main part of the envelope joins the neutron star and becomes
uniformly rotating with it, and the outer part of the envelope
expands with large velocity, carrying out a considerable part of
rotational energy and rotational momentum. MRE has an efficiency
about 10\% of the rotational energy,
 the ejected mass is $\approx 0.1Ì$ of the star mass,
explosion energy $\approx 10^{51}$ erg. Ejected mass and explosion
energy depend weekly on the parameter $\alpha =E_{mag}/E_{grav}$ at
initial moment. Explosion time depends on $\alpha$ as $t_{expl} \sim
\frac{1}{\sqrt{\alpha}}$. Small $\alpha$ is difficult for numerical
calculations with  explicit numerical schemes because of  the
Courant restriction on the time step, "hard" system of equations,
where
 $\alpha$  determines a "hardness".

\section{2-D  calculations}

The numerical method used in simulations is based on the implicit
operator-difference, completely conservative scheme on a Lagrangian
triangular grid of variable structure, with grid reconstruction
(Fig.\ref{grid}). The implicitness of the applied numerical scheme
allows for large time-steps. It is important to use the implicit
scheme in the presence of two strongly different time-scales: the
small one due to huge sound velocity in the central parts of the
star, and the big one determining the evolution of the magnetic
field. The method applied here was developed and its stability was
investigated in the papers of \cite{arch}, \cite{arkos1995},
\cite{ard96}. The scheme is fully conservative, what includes
conservation of mass, momentum and total energy, and correct
transitions between different types of energies. It was tested
thoroughly with different tests by \cite{abkm2000}.

\begin{myfigure}
\centerline{\includegraphics[width=6cm]{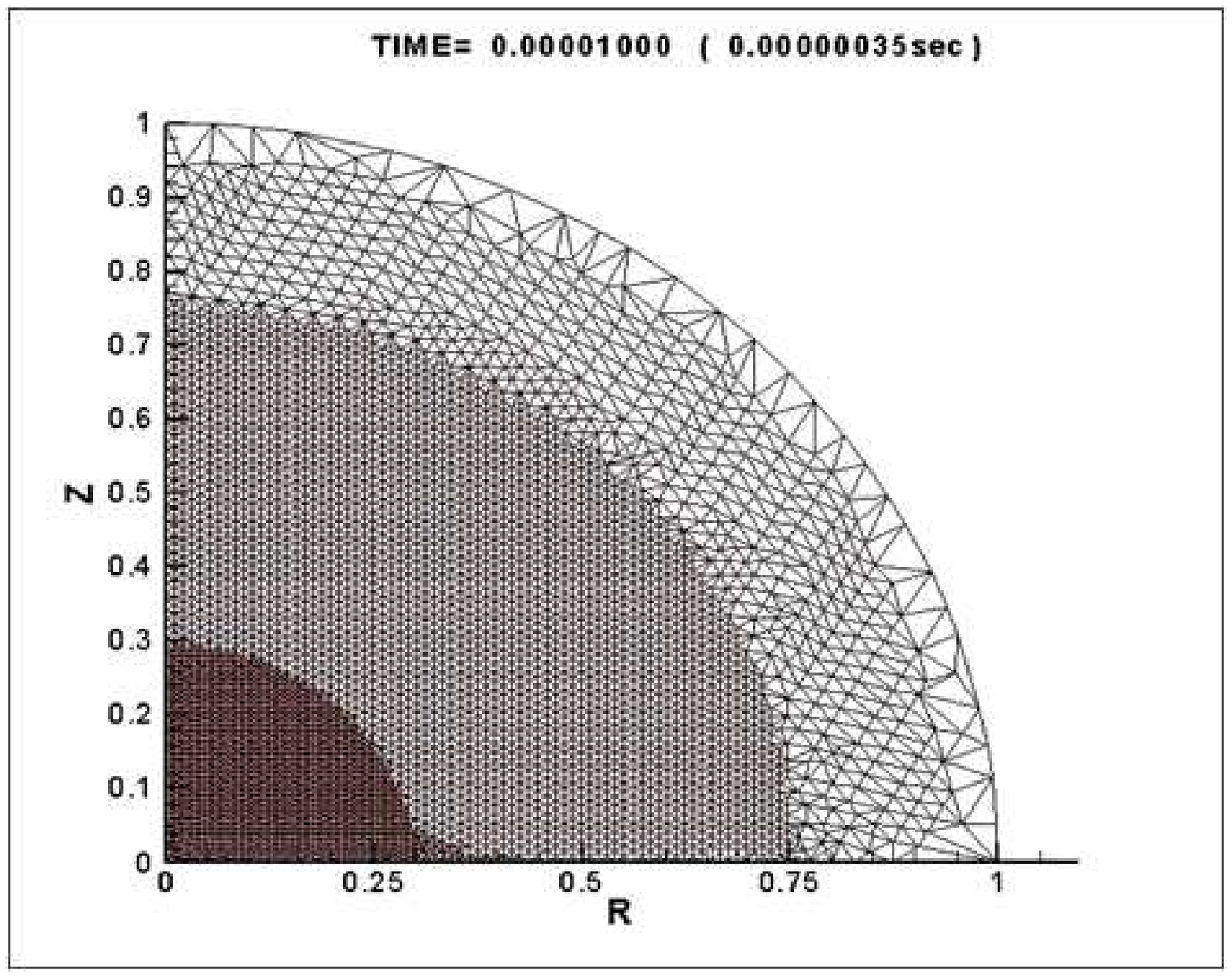}
}
 \caption{Example of the triangular grid }
  \label{grid}
\end{myfigure}

In the calculations of magnetorotational core-collapse supernova
performed by \cite{abkm2005}, magnetohydrodynamic (MHD) equations
with self-gravitation, and infinite conductivity have been solved
using the numerical scheme as described above.
  The problem has an axial symmetry (
$\frac{\partial}{\partial \phi}=0)$, and the symmetry to the
equatorial plane (z=0). Initial toroidal current $J_{\phi}$  was
taken at the initial moment (time started now from the stationary
rotating neutron star) producing $H_r, \,\, H_z$ according to
Biot-Savart law $ {\bf B}=\frac 1 c \int\limits_V^{} \frac {{\bf J}
\times {\bf R}}{{R}^3}dV$. Initial magnetic field of quadrupole-like
symmetry is obtained at opposite directions of the current in both
hemispheres. Neutrino cooling was calculated using a variant of a
flux-limited method, \cite{abkm2005}.

\begin{myfigure}
\centerline{\includegraphics[width=6cm]{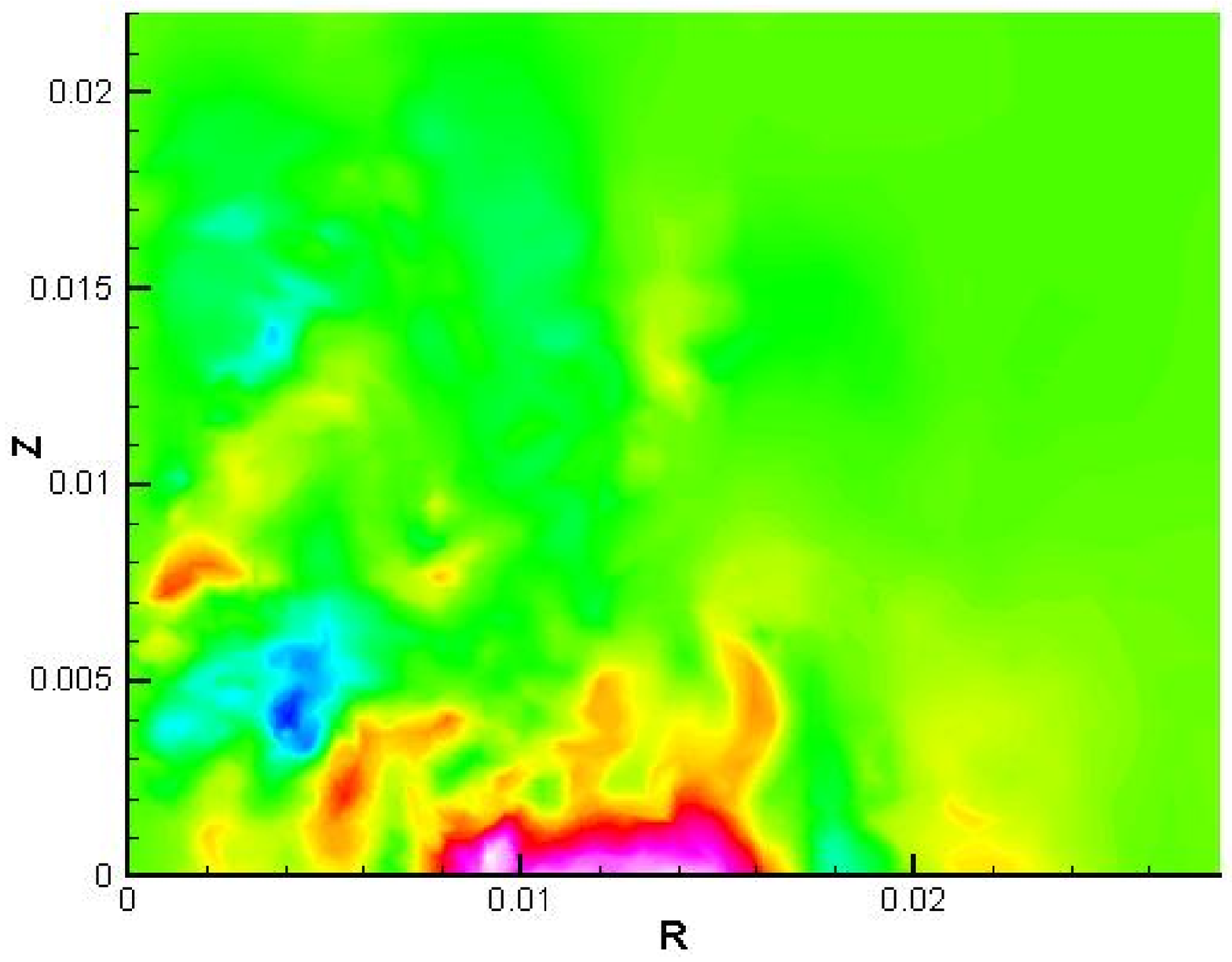}
}
 \caption{Toroidal magnetic field distribution at the moment of its
maximal energy for the initial quadrupole field.}
  \label{mag}
\end{myfigure}

 Magnetic field is amplified due to twisting by
the differential rotation, and subsequent development of the
magnetorotational instability. The field distribution for initial
quadrupole-like magnetic field with $\alpha=10^{-6}$, at the moment
of the maximal energy of the toroidal magnetic field is represented
in Fig.\ref{grid}. The maximal value of $B_\phi=2.5\cdot 10^{16}$ Gs
was obtained in the calculations. The magnetic field at the surface
of the neutron star after the explosion is $B=4 \cdot 10^{12}$ Gs.
 Time dependence during the explosion of rotational, gravitational,
internal, and kinetic poloidal energies is given in
Figs.\ref{engravetal}. Almost all gravitational energy, transforming
into heat during the collapse, is carried away by weakly interacting
neutrino.  The total energy ejected in the kinetic form is  $\sim
0.6\cdot 10^{51}$ erg, and the total ejected mass is equal to $\sim
0.14M_\odot$.

\begin{myfigure}
\centerline{\includegraphics[width=6cm]{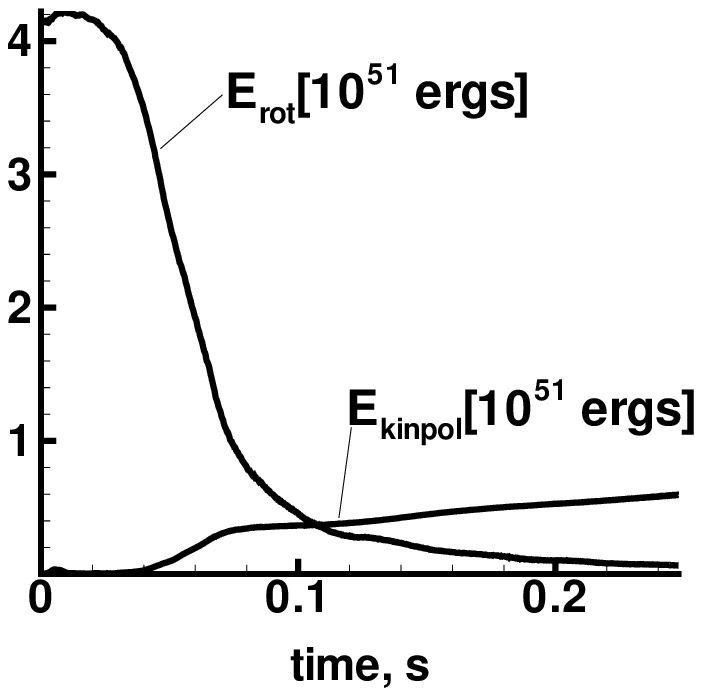}}
\caption {
{Time dependence of rotational, kinetic poloidal, and magnetic
energies during explosion for a dipole -like field,
 from \cite{mbka2006}.}}
\label{engravetal}
\end{myfigure}

The simulations were done for the initial poloidal magnetic field of
quadrupole \cite{abkm2005} and of dipole \cite{mbka2006} types of
symmetry. Before the collapse the ratios between the rotational and
gravitational, and between the internal and gravitational energies
of the star had been chosen as: $
\frac{E_{rot}}{E_{grav}}=0.0057,\quad
\frac{E_{int}}{E_{grav}}=0.727. $ The initial magnetic field was
"turned on" after the collapse stage. The ratio between the initial
magnetic  and  gravitational energies was chosen as $10^{-6}$. The
initial poloidal magnetic field in the center, at start of the
evolution of the toroidal field was $\sim 3.2\times 10^{13}$G.

\begin{myfigure}
\centerline{\includegraphics[width=6.5cm]{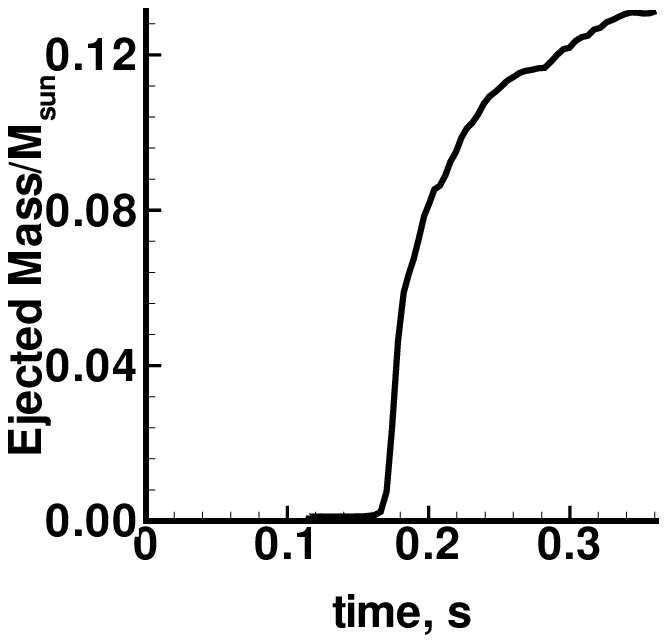}
}
 \caption{Time dependence of the ejected mass during the magnetorotational explosion with initial
 dipole magnetic field, from \cite{mbka2006}.}
  \label{ejmas}
\end{myfigure}

\begin{myfigure}
\centerline{
\includegraphics[width=6.5cm]{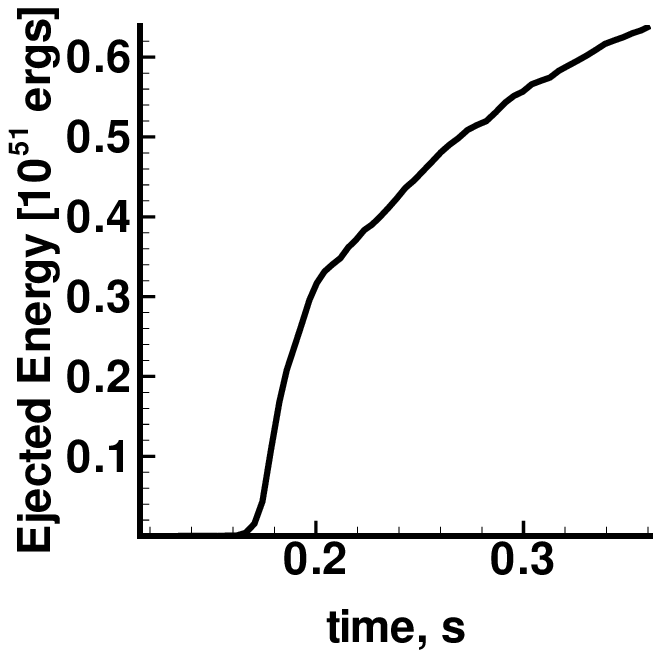}}
 \caption{Time dependence of the  ejected
 energy  during the magnetorotational explosion with initial
 dipole magnetic field, from \cite{mbka2006}.}
  \label{ejen}
\end{myfigure}

 The magnetic field  works as a piston for the originated MHD shock.
The time dependence of the ejected mass and energy is given in Figs.
\ref{ejmas}, \ref{ejen}.  During the magnetorotational explosion $\sim
0.14\>M_\odot$ of the mass  and  $\sim 0.6\cdot 10^{51}$ergs ($\sim
10\%$ of the rotational energy) are ejected. The simulation of the
MR supernova explosion for various initial core masses and
rotational energies was done by \cite{bkma2008}. The initial core
mass was varied from $1.2 M_\odot$ to $1.7 M_\odot$, the initial
specific rotational energy $E_{rot}/M_{core}$, was varied from 0$.19
\times 10^{19}$ to $0.4 \times 10^{19}$ erg/g. The explosive energy
increases  with the mass of the core, and the initial rotational
energy. The energy released in MR explosion, $(0.5-2.6) \times
10^{51}$ erg, is sufficient to explain supernova with collapsing
cores, Types II and Ib. The energies of Type Ic supernovae could be
higher.

\section{Magnetorotational instability}

Magnetorotational instability (MRI) leads to exponential growth of
magnetic fields. Different types of MRI have been studied by
\cite{vel59}, \cite{spru02}. MRI starts to develop when the ratio of
the toroidal to poloidal magnetic energies is becoming large. In 1-D
calculations MRI is absent because of a restricted degree of
freedom, and time of MR explosion is increasing with $\alpha$ as
$t_{\rm expl} \sim \frac{1}{\sqrt{\alpha}}$,
$\alpha=\frac{E_{mag0}}{E_{grav0}}$. Due to development of MRI the
time of MR explosion depends on $\alpha$ much weaker. The MR
explosion happens when the magnetic energy is becoming comparable to
the internal energy, at least in some parts of the star. While the
starting magnetic energy linearly depends on $\alpha$, and MRI leads
to exponential growth of the magnetic energy, the total time of MRE
in 2-D is growing {\bf logarithmically} with decreasing of $\alpha$,
$t_{expl} \sim -\log{\alpha}$. These dependencies are seen clearly
from 1-D (\cite{bkps1976}) and 2-D calculations (\cite{abkm2005},
\cite{mbka2006}) giving the following explosion times $t_{\rm expl}$
(in arbitrary units): $ \alpha=0.01,\,\, t_{\rm expl}=10, \quad
\alpha=10^{-12},\,\, t_{\rm expl}=10^6$ in 1-D, and $
\alpha=10^{-6},\,\, t_{\rm expl}\sim 6, \quad \alpha=10^{-12},\,\,
t_{\rm expl}\sim 12$  in 2-D. The dependence of the explosion time
is shown in  graphs for the quadrupole (\cite{mbka2005}), and dipole
(\cite{mbka2006}) configurations of the magnetic field. The
qualitative picture of the MRI in 2D,  and the example of the
analytical toy model with an exponential growth of the magnetic
field, have been  presented by \cite{abkm2005}, \cite{mbka2006}.

\section{Jet formation  in MRE}

Jet formation  in MRE happens when the initial magnetic field is of
a dipole-like structure. 2-D calculations with the initial
dipole-like magnetic field gave almost the same values of the energy
of explosion $\sim 0.5 \cdot 10^{51}$ ýðã,
 and ejected mass $\approx 0.14M_{\odot}$,
but the outburst was slightly collimated along the rotational axis
\cite{mbka2006}, see Figs.{\ref{dipt1}}.{\ref{dipt2}}

\begin{myfigure}
\centerline{\includegraphics[width=6.5cm]{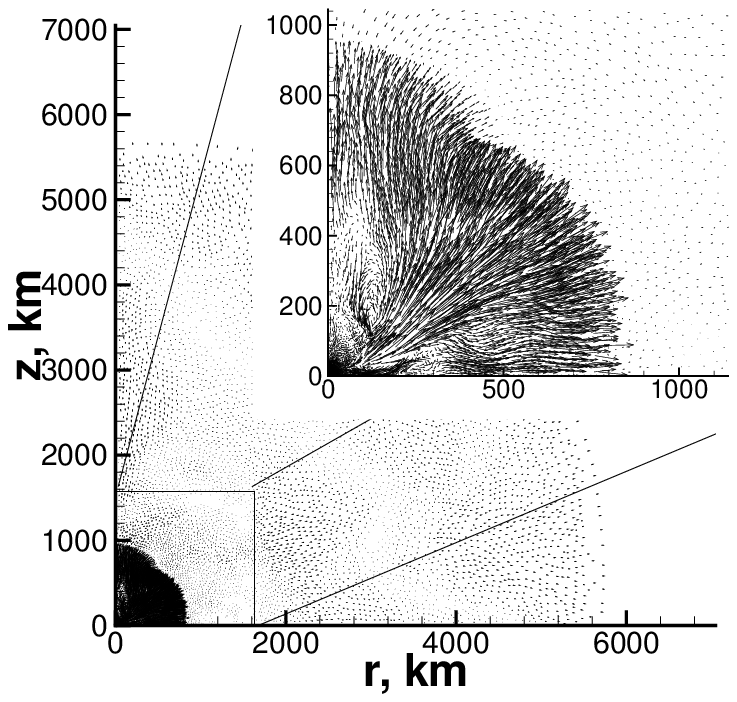}
}
\caption {{ Time evolution of the velocity field (outflow) for the
time moment $t=0.075$s,
  from \cite{mbka2006}}}
\label{dipt1}
\end{myfigure}

\begin{myfigure}
\centerline{
\includegraphics[width=6.5cm]{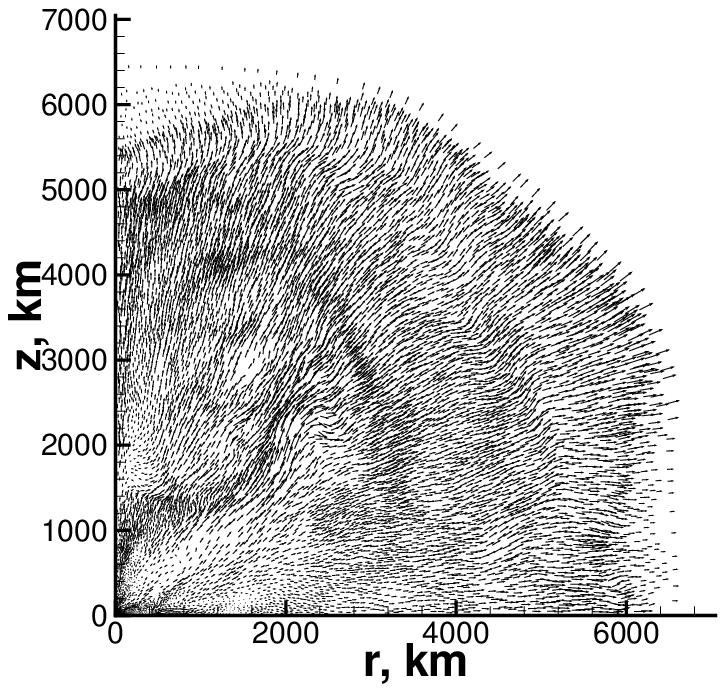}}
\caption {{Time evolution of the velocity field (outflow) for
the time moment, $t= 0.25s$,
  from \cite{mbka2006}.}}
\label{dipt2}
\end{myfigure}

\noindent  Simulations of
the MR supernova have been made with equation of state suggested in
\cite{shen1998}. A comparison of our results for the initially
uniform magnetic field, using a Lagrangian scheme, with the results
in \cite{takiwaki2004} and \cite{tkw2009}, using an Eulerian scheme
for the same initial and boundary conditions, shows good agreement
for a strong initial field ($H_0=10^{12}$G), while for a weaker
field ($H_0=10^9$G) we  get mildly collimated jet-like explosion;
see also \cite{bur07}. Details of the results of these simulations
will be published elsewhere (\cite{m2012}).  MRI is developed in the
case of a weaker initial magnetic field, and it is not present in
the calculations with stronger field, see Figs.\ref{mri1},\ref{nomri}.

We have made simulations for the  initial magnetic fields
$
    H_0=10^9,\,\,\,10^{12} G.
$
The ratio of the initial rotational energy to the absolute value of the gravitational energy was taken
$
    E_{rot0}/E_{grav0}=1\% , \,\,\,2\%.
$
When the initial magnetic field is moderate ($H_0=10^9\,$G) and $E_{rot0}/E_{grav0}=1\%$ MDRI develops what means exponential growth of all components of the magnetic field (Fig.\ref{mri1}).

At Fig.\ref{etorepol} The Lagrangian triangular grid and the ratio of the toroidal magnetic energy to the poloidal one ($E_{tor}/E_{pol}$) is represented for the case $H_0=10^9\,$G, and $E_{rot0}/E_{grav0}=1\%$. The toroidal magnetic energy dominates over the poloidal one in the significant part of the region where new neutron star is forming. The MDRI is well-resolved on our triangular grid.

In the case ($H_0=10^{12}\,$G) and $E_{rot0}/E_{grav0}=1\%$ there is no regions of domination of $E_{tor}$ over $E_{pol}$.

The Fig.\ref{erotmag9} represents a time evolution of rotational, magnetic poloidal and toroidal energies for MR explosion when $H_0=10^9$G.

The Fig.\ref{erotmag9z} is the same data plot as the Fig.\ref{erotmag9} but zoomed and the vertical axis is in logarithmic scale. The straight dash-dotted line at the Fig.\ref{erotmag9z} shows the exponential growth of the toroidal and poloidal magnetic energies with the time due to the MDRI.

The rotational energy has two maxima. The first  contraction is accompanied by   the strong growth of the rotational energy due to angular momentum conservation, maximum of which coincides with the first maximum of the density. The first contraction, and the subsequent bounce, happens when the magnetic field is growing slowly, and the angular momentum losses from the stellar core are small. Development of the magnetorotational instability leads to a rapid growth of the magnetic field, large angular momentum flux from the core, what stops the expansion, and leads to the second contraction phase. In this case the contraction is not transforms into expansion, because of the rapid decrease of the rotational energy due to strong angular momentum flux outside from the core.

The energy of the poloidal magnetic field grows due to the contraction until the time $t\approx 0.225$ sec.
Then it slightly decreases  because of the formation of the bounce shock, and its motion outwards. The toroidal magnetic energy grows as quadratic function because of wrapping of the magnetic force lines (toroidal component of the magnetic field grows linearly). Starting from $t\approx 0.3$ sec both the poloidal and toroidal magnetic energies begin to grow exponentially  due to MDRI. At $t\approx 0.36$ sec both magnetic energies comes to saturation. The MHD shock wave develops  what leads to the MR explosion. The MR explosion develops in all directions without formation of a collimated flow.

The MR explosion for an extremely high initial magnetic field  ($H_0=10^{12}$ G) is developing in a qualitatively different way. The initial magnetic field is so strong that it grows strongly during the first contraction, and the explosion happens before  the development of MDRI happens (Fig.\ref{nomri}).
At the Fig.\ref{erotmag12} the time evolution of the rotational, poloidal magnetic and toroidal magnetic energies are represented.
The rotational energy has one extremum at $t\approx 0.32$ sec corresponds to the maximal contraction, accompanied by a corresponding growth of of the toroidal and poloidal magnetic energies. The poloidal magnetic field grows due to the contraction, the toroidal magnetic field appears due to the differential rotation and is amplified both due to the differential rotation and the contraction of the core.

The strong initial magnetic field leads to a rapid loss of the angular momentum from the core already during the contraction phase. The centrifugal force becomes unimportant, and the first contraction is not followed by any bounce, leaving behind a dense slowly rotating neutron star core.
We have got here a prompt explosion.

The force lines of the magnetic field play the role of 'rails'. The matter moves along the force lines. The magnetic pressure dominates a the periphery of the core. The MR explosion develops mainly along the axis of rotation, and the collimated flow (protojet) is formed.
The MR explosion results in the collimated jet. The degree of jet collimation is approximately the same as in \cite{takiwaki2004}.
For the case when $ E_{rot0}/E_{grav0}=1\%$ and $H_0=10^9$ G the MR supernova explosion energy is $\sim 4\times 10^{50}$ erg, for the $H_0=10^{12}\,$G the MR supernova explosion energy reaches the value of $\sim 8\times 10^{50}$erg.
The explosion energy resulted in the simulations by Lagrangian method are close to those ones found in the simulations made by Eulerian scheme \cite{takiwaki2004} (excluding the case of $B_0=10^{12}\,G$ and $E_{rot0}/|E_{grav0}|=1/\%$).

\vspace{-2cm}
\begin{myfigure}
\centerline{
\includegraphics[width=8cm, angle=-90]{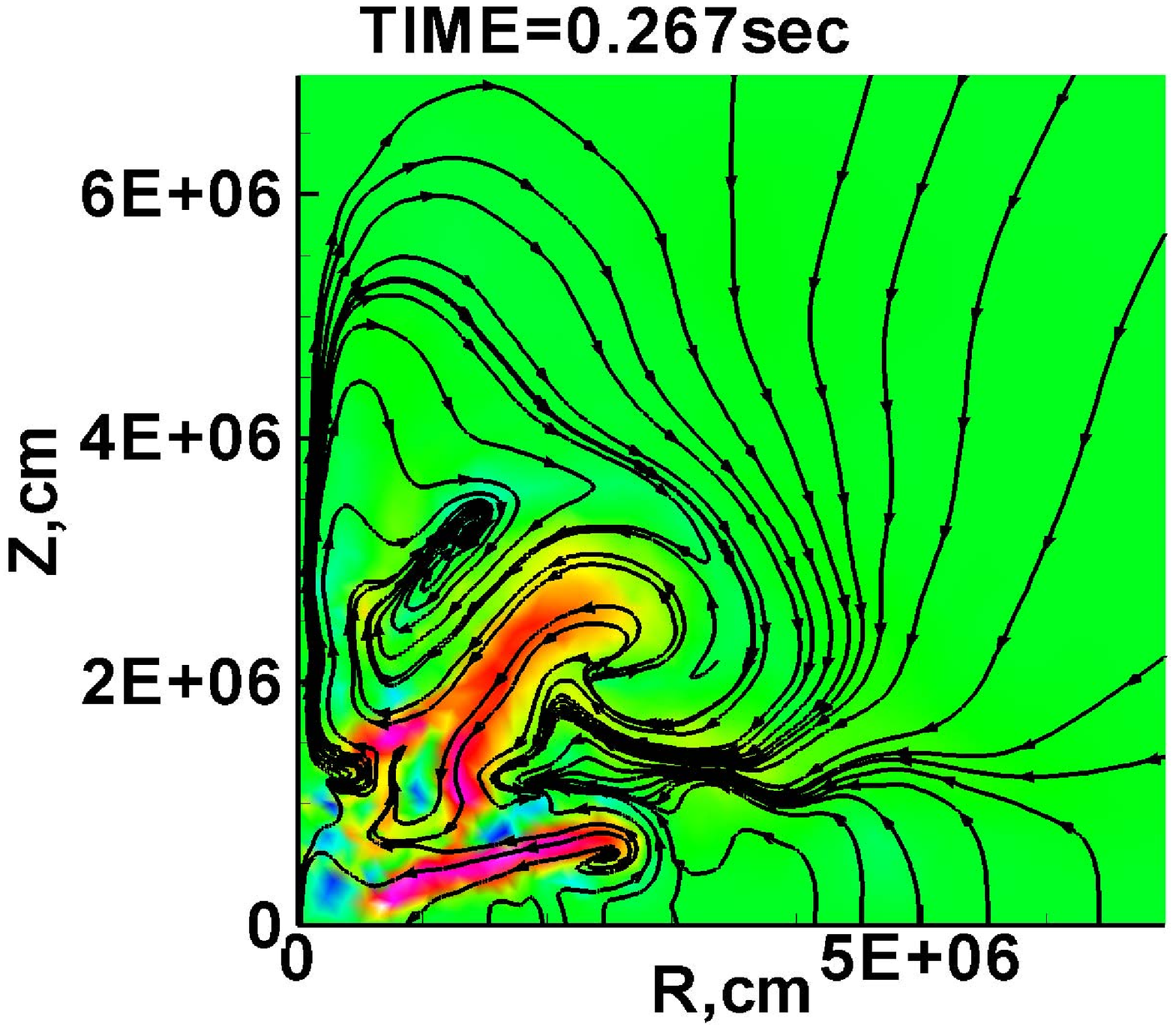}}
 \caption{Developed MDRI due to convection and MRI/Tayler
instability at $t=267$ms for the case $H_0=10^9 {\rm G}, E_{rot0}/E_{grav0}=1\%$
(contour plot - the toroidal magnetic field, arrow lines - force lines of the poloidal magnetic field).}
  \label{mri1}
\end{myfigure}

\begin{myfigure}
\centerline{
\includegraphics[width=8cm]{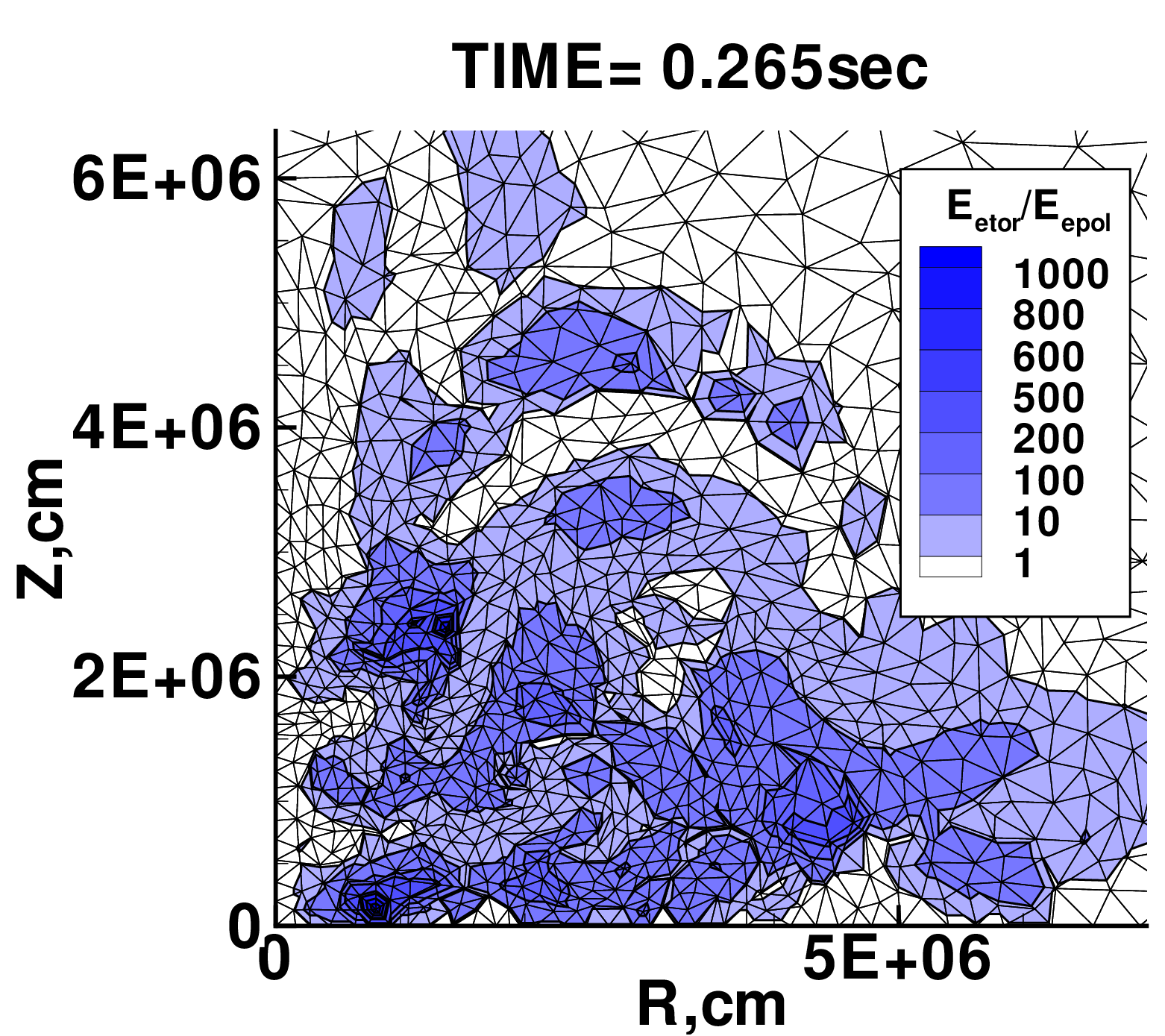}}
 \caption{The Lagrangian triangular grid and the ratio of the toroidal magnetic energy $E_{tor}$ to the poloidal magnetic energy $E_{pol}$, $\frac{E_{tor}}{E_{pol}}$ at $t=265$ms for the case $H_0=10^9 {\rm G}, E_{rot0}/E_{grav0}=1\%$.}
  \label{etorepol}
\end{myfigure}

\vspace{-3cm}
\begin{myfigure}
\centerline{
\includegraphics[width=8cm]{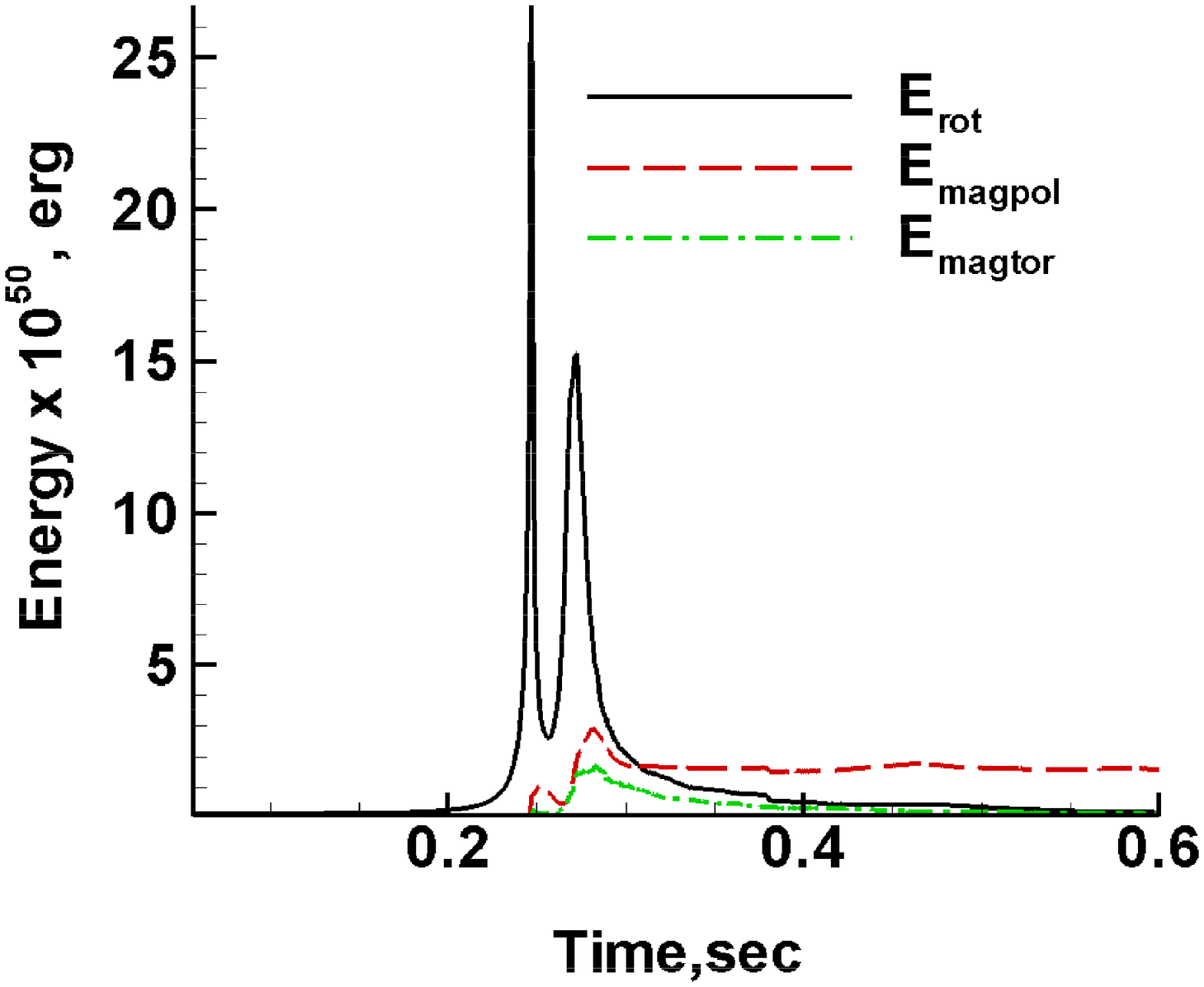}}
 \vspace{-2cm}\caption{Time evolution of rotational energy $E_{rot}$ (solid line), magnetic poloidal energy $E_{magpol}$ (dashed line) and magnetic toroidal energy $E_{magtor}$ (dash-dotted line) for the case $H_0=10^{9} {\rm G}, E_{rot0}/E_{grav0}=1\%$.}
  \label{erotmag9}
\end{myfigure}

\section{Asymmetry of the explosion}

It is known from the observations that the shapes of core collapse
supernovae are different. From our simulations it follows that MR
supernova explosion arises after development of the MRI. The
development of the MRI is a stochastic process and hence the
resulting shape of the supernova can vary.
 We may conclude that MR supernova
explosion mechanism can lead to different shape of the supernova.
It is important to point out that MR mechanism of supernova
explosion leads always to asymmetrical outbursts.

The simulations of the MR supernova explosions described here are
restricted by the symmetry to the equatorial plane.
While in reality this symmetry can be violated due to the MRI,
simultaneous presence of the  initial dipole and quadrupole -like
magnetic field (\cite{wang}) and initial toroidal magnetic field
(\cite{bkm1992}). The violation of the symmetry could lead to
the kick effect and formation of rapidly moving radio pulsars.
A kick velocity, along the rotational axis, formed due to magnetohydrodynamic processes
in presence of the asymmetry of the
magnetic field, by estimations \cite{bk93} does not exceed  ~300km/sec.

When rotational and magnetic axes do not
coincide the whole picture of the explosion process is three
dimensional. Nevertheless, the magnetic field twisting happens always
around the rotational axis, so we may expect the kick velocity of the
neutron star be strongly correlated with its spin direction. During the phase of MRE explosion
the regular component of magnetic field may exceed temporally $10^{16}$ G \cite{abkm2005}, \cite{mbka2006},
when the neutrino cross-section depends on the magnetic field strength.
The level of the anisotropy of the magnetic field relative to the plain
perpendicular to the rotational axis \cite{hanawa} may be of the order of 50\%, leading to
strong anisotropy of the neutrino flux.  The kick velocity due to
the anisotropy of the neutrino flux may reach several thousands km/c \cite{bk93}, explaining
appearance of the most rapidly moving radio pulsars \cite{kick}.
Simultaneously, because of the stochastic nature of MRI, the level of the
anisotropy should be strongly variable, leading to a large spreading in the
the neutron star velocities.
This prediction
of MR explosion differs from the models with a powerful neutrino convection, where
arbitrary direction of the kick velocity is expected (\cite{bur},\cite{jan}).
It was claimed in \cite{db06}, that proto-neutron star (PNS) convection was  found to be a secondary
feature of the core-collapse phenomenon, rather than a decisive ingredient for a successful explosion.

Analysis of observations of pulsars shows that
rotation and velocity vectors of pulsars are aligned, as is predicted
by the MR supernova mechanism. This alignment was first found in \cite{ssm96}, and was
confirmed with reliability, increasing with time, in the papers \cite{jon05},\cite{jon07},\cite{jon12}.
 The alignment of the vectors can
be violated in the case when the supernova explodes in a binary
system.

\section{Conclusions}

In MRE the efficiency of transformation of rotational energy into
the energy of explosion is $\sim 10$\%. MRI strongly accelerates
MRE, at lower values of the initial magnetic fields. Jet formation
is possible for dipole-like topology of the field. MRE  energy is
not sensitive to the details of the equation of state, model of the
neutrino transfer,  and to the choice of the numerical scheme.
The observed alignment of the rotation and velocity vectors of pulsars follows directly
from the MRE supernova model.

\begin{myfigure}
\centerline{
\includegraphics[width=8cm, angle=-90]{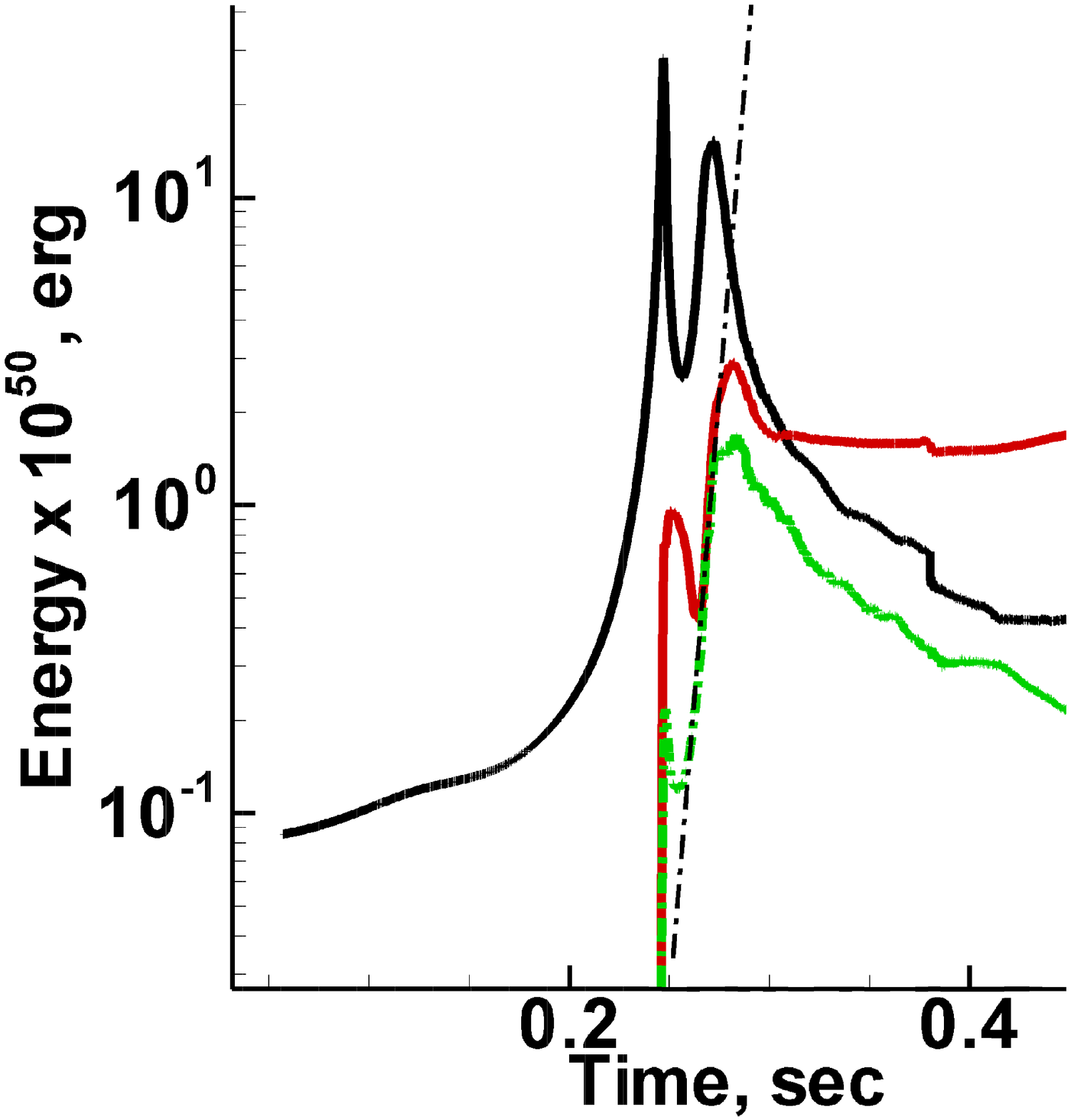}}
 \caption{Zoomed time evolution of rotational energy $E_{rot}$ (solid line), magnetic poloidal
 energy $E_{magpol}$ (dashed line) and magnetic toroidal energy $E_{magtor}$ (dash-dotted line)
 for the case $H_0=10^{9} {\rm G}, E_{rot0}/E_{grav0}=1\%$. Straight dash-dotted line shows
 exponential growth of the toroidal and poloidal magnetic energies.}
  \label{erotmag9z}
\end{myfigure}

\begin{myfigure}
\centerline{
\includegraphics[width=8cm]{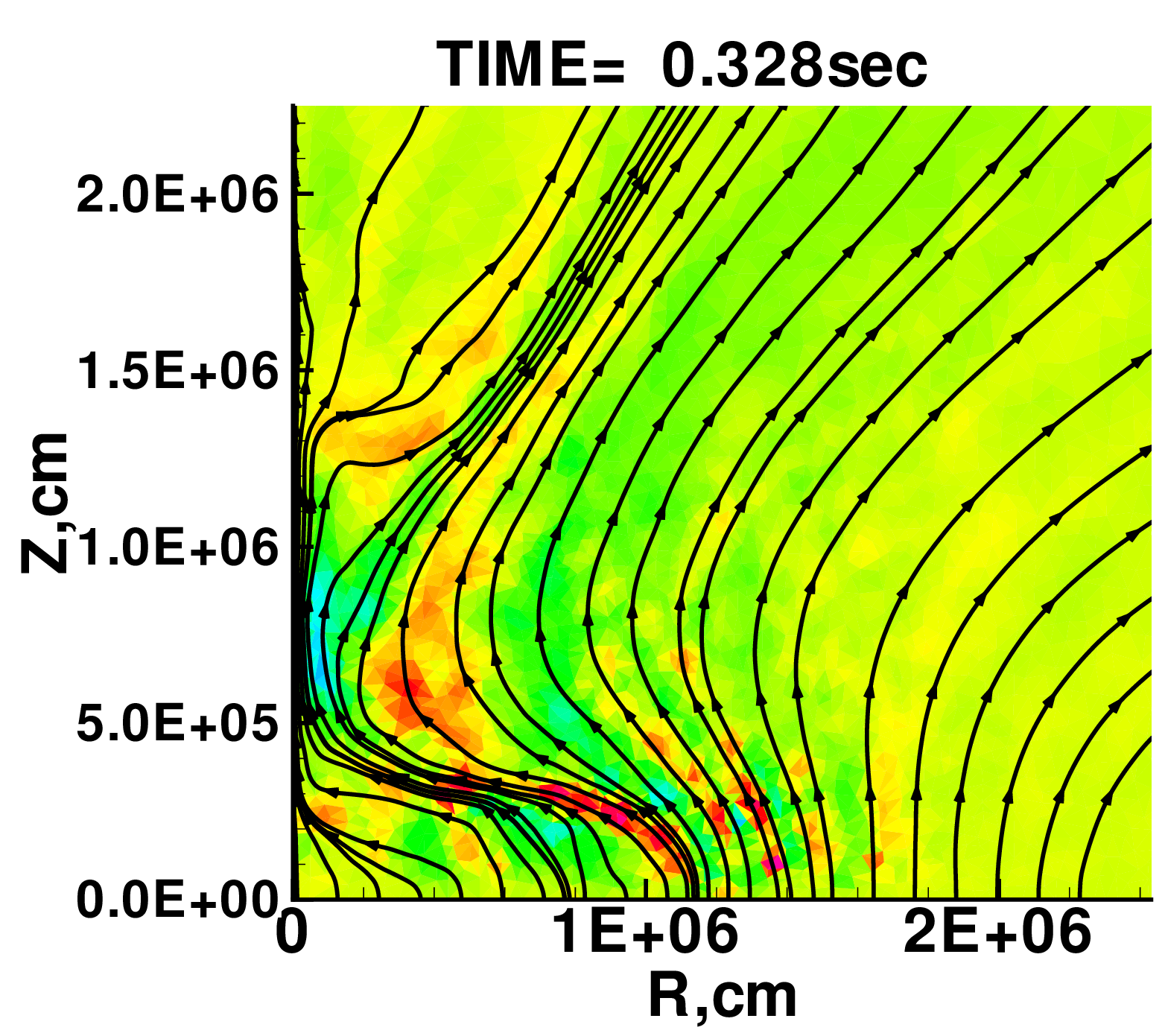}}
 \caption{Absence of MDRI at $t=328$ms for the case $H_0=10^{12} {\rm G}, E_{rot0}/E_{grav0}=1\%$
 (contour plot - the toroidal magnetic field, arrow lines - force lines of the poloidal magnetic field).}
  \label{nomri}
\end{myfigure}

\begin{myfigure}
\centerline{
\includegraphics[width=8cm, angle=0]{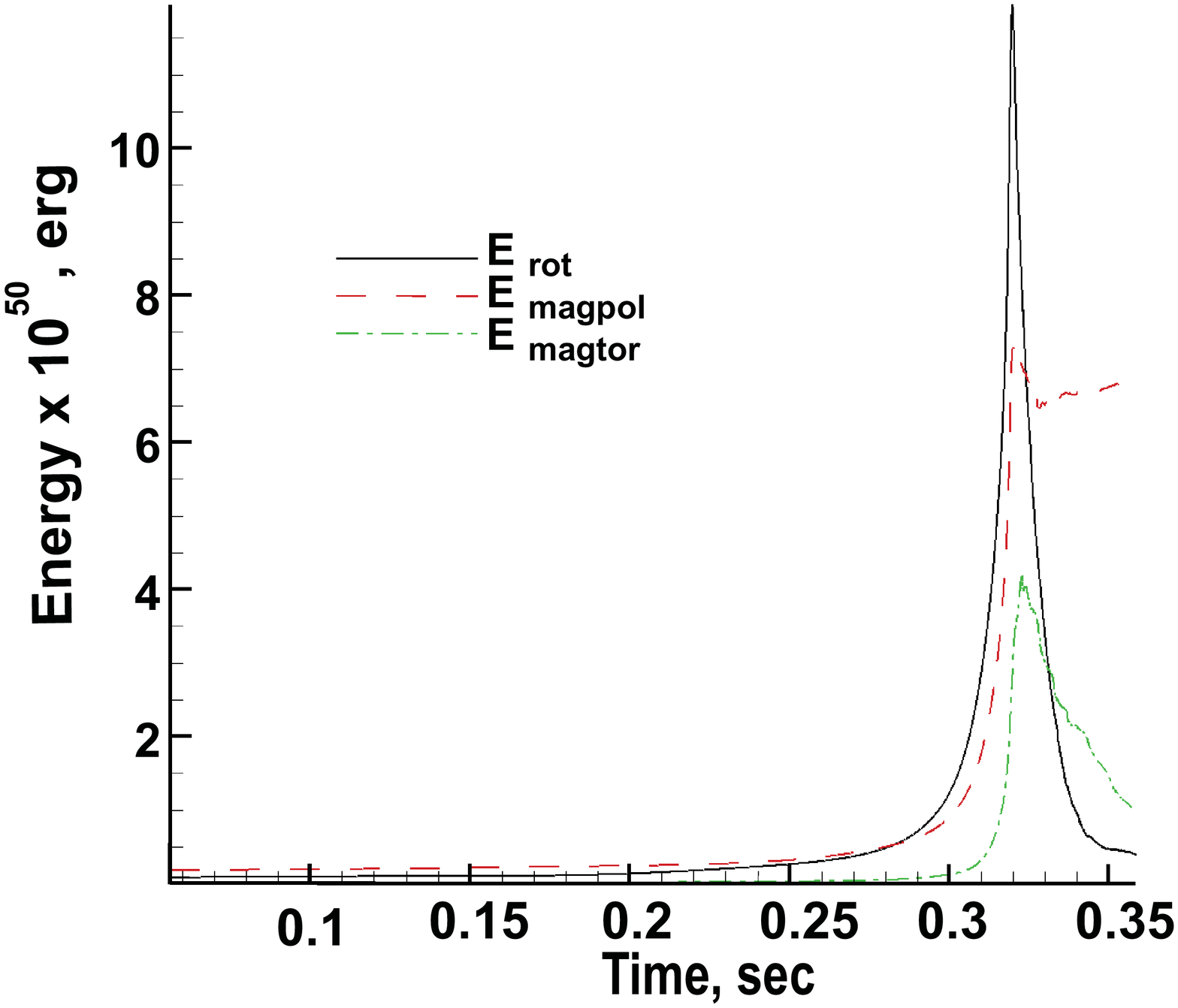}}
 \caption{Time evolution of rotational energy $E_{rot}$ (solid line),
 magnetic poloidal energy $E_{magpol}$ (dashed line) and magnetic toroidal energy $E_{magtor}$
 (dash-dotted line) for the case $H_0=10^{12} {\rm G}, E_{rot0}/E_{grav0}=1\%$.}
  \label{erotmag12}
\end{myfigure}

\thanks The work of SGM and GSBK was supported partially by
RFBR grant 14-02-00728, grant for leading scientific schools
NSH-261.2014.2,  and  RAS  program 21/3.


\end{multicols}
\end{document}